\documentclass[amssymb, amsmath, aps, prb, twocolumn, superscriptaddress, showpacs]{revtex4}
\usepackage{bm,url,graphicx,natbib}%

\newcommand{\degr}{\ensuremath{^{\circ}}}

\begin{document}
\title{Carrier dynamics in ion-implanted GaAs studied by simulation and observation of terahertz emission}%
\author{J.~Lloyd-Hughes}
\email{james.lloyd-hughes@physics.ox.ac.uk}

\author{E.~Castro-Camus}
\affiliation{{\footnotesize Department of Physics, University of Oxford, Clarendon Laboratory, Parks Road, Oxford, OX1 3PU, United Kingdom}}

\author{M.D.~Fraser}
\author{C.~Jagadish}
\affiliation{{\footnotesize Department of Electronic Materials Engineering, Research School of Physical Sciences and Engineering, Institute of Advanced Studies, Australian National University, Canberra ACT 0200, Australia}}%

\author{M.B.~Johnston}
\affiliation{{\footnotesize Department of Physics, University of Oxford, Clarendon Laboratory, Parks Road, Oxford, OX1 3PU, United Kingdom}}
\parbox{17.2cm}{\centering {\footnotesize PHYSICAL REVIEW B {\bf 70}, 235330 (2004)}}
\received{23 July 2004}
\published{22 December 2004}

\begin{abstract}
{We have studied terahertz (THz) emission from arsenic-ion implanted GaAs both experimentally and
using a three-dimensional carrier dynamics simulation. A uniform density of vacancies was formed
over the optical absorption depth of bulk GaAs samples by performing multi-energy implantations of
arsenic ions (1 and 2.4\,MeV) and subsequent thermal annealing. In a series of THz emission
experiments the frequency of peak THz power was found to increase significantly from 1.4 to
2.2\,THz when the ion implantation dose was increased from $10^{13}$ to $10^{16}$ cm$^{-3}$. We
used a semi-classical Monte-Carlo simulation of ultra-fast carrier dynamics to reproduce and
explain these results. The effect of the ion-induced damage was included in the simulation by
considering carrier scattering at neutral and charged impurities, as well as carrier trapping at
defect sites. Higher vacancy concentrations and shorter carrier trapping times both contributed to
shorter simulated THz pulses, the latter being more important over experimentally realistic
parameter ranges.}
\end{abstract}

\pacs{78.47.+p, 78.20.Bh, 61.72.Vv, 68.55.Ln \parbox{6.7cm}{\raggedleft DOI: 10.1103/PhysRevB.70.235330}}
\maketitle

\fontsize{9}{10.8}
\selectfont
\pagestyle{myheadings}
\markboth{}{\footnotesize{{\rm LLOYD-HUGHES {\it et al.}, PHYSICAL REVIEW B {\bf 70} 235330 (2004)}}}

\section{Introduction}
\label{SEC:introduction}
The technique of terahertz time-domain spectroscopy (THz-TDS) has seen wide application in recent years.
 By resolving carrier dynamics on ultra-short time scales THz-TDS has proven invaluable in areas
 of physics as disparate as investigating the energy gap in high T$_{\rm{c}}$ superconductors\cite{prl88_027003},
 determining low-energy vibrational modes in oligomers\cite{cpl03}, and
 observing the onset of quasi-particle screening in GaAs\cite{nat414_286}.

Such studies have become possible through the development
 of large bandwidth, high power THz emission sources based on the ultra-fast separation of photoexcited carriers
 in semiconductors. The resulting change in current produces an electromagnetic transient, typically a single or
 half-cycle pulse, with a broad spectrum covering THz frequencies. Charge separation can be the result
 of differing electron and hole
 mobilities (the photo-Dember effect), or alternatively may result from an electric field accelerating
 electrons and holes in opposite directions. In the later case, the electric field can be internal,
 such as the surface depletion field near a semiconductor surface, or externally applied as in a photoconductive
 switch (PCS). The study of surface emitters can help to
 elucidate aspects of ultra-fast carrier dynamics, such as the enhancement of THz emission under a magnetic
 field \cite{prb01,jap02,prb02}, and the relative importance of the photo-Dember and surface field charge separation
 mechanisms in GaAs and InAs \cite{prb02,apl83_5476}. Additionally, surface emitters can produce
 the well-collimated THz beams \cite{sst04} desirable for many THz imaging applications.

In the effort to improve THz emitters, one method for achieving a greater bandwidth is to shorten the electric field rise
 time, normally limited by the duration of the exciting infrared pulse. Another
 approach is to reduce the duration of the electric field transient after excitation by choosing a semiconductor with
 a short carrier lifetime, and/or a high momentum scattering rate. One suitable material is low-temperature-grown
 (or LT-) GaAs, in which {As} precipitates and trapping centres result in a short ($<1$\,ps) carrier
 lifetime, enabling THz emission extending to over $30$\,THz \cite{apl83_3117}. Alternatively the
 technique of ion implantation can be used, in which incident energetic ions damage the crystal
 structure of a semiconductor. The principal defect in arsenic ion-implanted GaAs (GaAs:As$^+$) is
 thought to be the antisite donor defect As$_{\rm{Ga}}$, where As replaces
 Ga on some Ga lattice sites. Deep-level transient spectroscopy has found that the defect
 energy levels in GaAs:As$^+$ lay $\sim$0.3\,eV below the conduction band\cite{jap81_7295}. Such
 deep-level defects can decrease THz pulse duration
 (increase THz bandwidth) by acting as carrier trapping and scattering centres, and
 result in carrier lifetimes of the order of $0.1$\,ps \cite{apl66_3304, apl76_1306}.

The controllable ion dosage permits ion-implanted GaAs samples with reproducible
properties, in contrast
 to LT-GaAs where the difficulty of temperature control results in samples with nominally the same growth
 parameters having varying properties. Ion-implantation has the further advantage of generating
 depth-dependent damage profiles (by choosing the ion type, dose, and energy) in specific sample areas. By performing
 a post-implant anneal the resistivity of GaAs:As$^+$ can be increased to levels
 comparable to LT-GaAs \cite{jqe34_1740, apl76_1306}, as required for photoconductive switches.

Both surface field\cite{apb72_151} and photoconductive antenna\cite{jap93_2996} ion-implanted
 THz emitters have been previously studied, and were found to produce THz radiation at slightly higher frequencies than
 semi-insulating GaAs\cite{jap93_2996}. Additionally, ion-implanted photoconductive antennae have been used as
 detectors \cite{apl83_1322}, but no bandwidth improvement could be seen. These studies utilised low
 energy ions, typically $200$\,keV, resulting in
 implantation depths of only $\sim$0.1\,$\mu$m. However, as the infrared absorption depth
 is $\sim$0.8\,$\mu$m in GaAs only $\sim$10\% of photocarriers are generated within this distance, and
 the non-implanted layer dominates THz emission.

In this paper we investigate GaAs implanted with dual, high energy (1\,MeV and 2.4\,MeV) {As} ion doses, creating an
 approximately uniform damage profile extending $\sim$1\,$\mu$m into the semiconductor. In Section \ref{SEC:exp} we
 describe the terahertz emission from samples irradiated with various ion doses, and discuss
 the effect of a post-implant thermal annealing step. Simulation results of THz emission from ion-implanted GaAs
 obtained from a three-dimensional carrier dynamics model\cite{prb02} are presented in Section \ref{SEC:sim}.

\section{Experiment}
\label{SEC:exp}

\begin{figure}[bt]
    \centering
    \includegraphics[width=6cm]{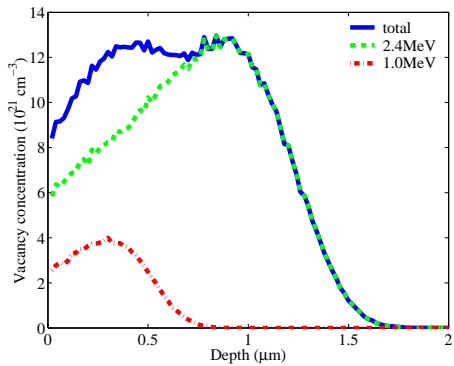}
    \caption{\label{FIG:profile} (Colour online) Depth dependence of vacancies (including recoil vacancies) calculated using SRIM
 software \cite{SRIM} for the sample with highest ion dose. The 1\,MeV, $2.5\times10^{15}$\,cm$^{-2}$ dose
 (dash-dotted line) and 2.4\,MeV, $1.0\times10^{16}$\,cm$^{-2}$ dose (dashed line) create an approximately
 uniform damage profile (solid line) extending over 1\,$\mu$m from the surface. By scaling the concentration
 according to the dose, the distributions for lower dose samples can be obtained.
}
\end{figure}

The Australian National University $1.7$\,MV tandem accelerator was used to irradiate GaAs samples from the
 same ingot with As$^+$ ions. By employing a two-stage implant of 1\,MeV and 2.4\,MeV (implanted at room temperature)
 an approximately uniform damage profile $>$1$\mu$m in depth was created (Fig.\ \ref{FIG:profile}) in
 7 samples at incident ion doses ranging from
 $2.5\times10^{12}$\,cm$^{-2}$ to $2.5\times10^{15}$\,cm$^{-2}$ (1\,MeV implant) and
 $1\times10^{13}$\,cm$^{-2}$ to $1\times10^{16}$\,cm$^{-2}$ (2.4\,MeV implant). A duplicate set of
 samples was produced that were annealed for 30\,min.\ at 500{\degr}C under AsH$_3$.
Calculations using the Stopping Range of Ions in Matter (SRIM) software\cite{SRIM} predict that the
 implanted ion concentration in the lowest dose sample is $\sim 1\times10^{17}$\,cm$^{-3}$. Assuming
 that $99\%$ of As$_{\rm{Ga}}$ antisites are instantly annealed\cite{SRIM} due to the implantation
 being performed at room temperature, we estimate an n-type doping concentration of $\sim 1\times10^{15}$\,cm$^{-3}$ for
 the lowest dose sample. Damage distributions were also calculated using SRIM, and
 verified using X-ray diffraction. Approximately 10,000 vacancies were found to be produced
 per As$^{+}$ ion.

The surface field THz emission from each GaAs:As$^+$ sample was studied using THz-TDS.
 Infrared pulses from a mode-locked $10$\,fs Ti-sapphire laser
 ($\lambda=800$\,nm, $\Delta \lambda = 135$\,nm, 75MHz repetition rate, $\sim400$\,mW typical beam power) were
 incident on the sample
 at 45\degr. A lens focused the infrared pulses to a Gaussian beam-waist of $\sigma =90$\,$\mu$m. The
 sample was positioned
 $\sim$1\,cm from the pump beam focus, with a Gaussian beam-waist of $\sigma = 160$\,$\mu$m
 chosen to provide a collimated THz beam\cite{sst04}. The THz radiation was collected by an off-axis
 parabolic mirror placed at its focal length from the pump beam focus, and refocused by
 an identical mirror onto a $0.2$\,mm thick $\langle 1\overline{1}0 \rangle$ ZnTe crystal. A delayed probe beam was
 split from the pump using a
 $10\%$ beam-splitter, and also illuminated the ZnTe. The change in the probe's polarization was measured as
 a function of delay using an electro-optic sampling (EOS) system consisting of a Wollaston
 prism, $\lambda/4$ waveplate and balanced
 photodiodes. The pump beam was chopped at $\sim1$\,kHz, enabling the voltage across the detector to be measured
 by a lock-in amplifier (200ms time constant), from which the THz electric field was
 calculated \cite{rsi73_1715}. Following the procedure of Zhao \emph{et.~al.}\ \cite{rsi73_1715} we
 determine that our EOS system operates in the shot-noise limit, with a noise-equivalent THz electric field
 $E_{\rm{NEF}} \simeq 0.01$\,Vm$^{-1}$ consistent with the experimentally observed root-mean-square
 noise $E_{\rm{RMS}}\simeq 0.01\rightarrow 0.02$\,Vm$^{-1}$.

The THz electric field emitted by each of the 7 annealed samples was thus measured, with signal-to-noise ratios of
 above 150:1. The maximum THz electric field $E_{\rm{THz}}^{\rm{max}}$ was of the order
 of $3$\,Vm$^{-1}$, less than we observe for our unimplanted GaAs reference sample ($4$\,Vm$^{-1}$), and
 an InAs sample ($216$\,Vm$^{-1}$) with the same experimental setup. For comparison, the maximum
 THz electric field strength reported in the literature from a photoconductive antenna
 is $9500$\,Vm$^{-1}$ (Ref.~\onlinecite{rsi73_1715}). $E_{\rm{THz}}^{\rm{max}}$ was not found to
 vary systematically with ion dose, perhaps because of variations in the surface potential of the samples,
 which can greatly alter the surface field strength.

\begin{figure}[tb]
    \centering
    \includegraphics[width=6cm]{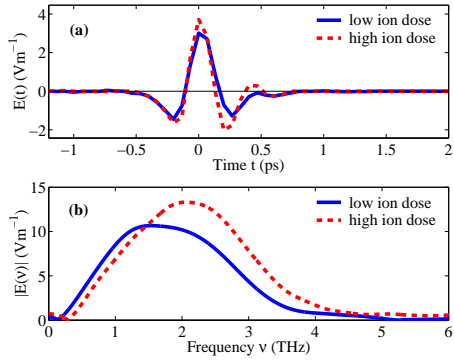}
    \caption{(Colour online) Measured THz electric field $E_{\rm{THz}}$ emitted via the surface field mechanism
 from GaAs:As$^{+}$ samples implanted with low
    doses (solid lines, $1\times 10^{13}$\,cm$^{-2}$ at 2.4\,MeV and $2.5\times 10^{12}$\,cm$^{-2}$ at 1\,MeV)
 and high doses (dotted lines, $5\times 10^{15}$\,cm$^{-2}$ at 2.4\,MeV and $1.25\times10^{15}$\,cm$^{-2}$ at 1\,MeV).
 The time and frequency domain THz electric fields are shown in (a) and (b). The peak electric fields
 and trailing negative peaks are $+3.0, -1.5$\,Vm$^{-1}$ and $+3.7, -2.0$\,Vm$^{-1}$ for the low
 and high dose samples respectively.}
\label{FIG:dose}
\end{figure}

Typical THz electric fields are shown in Fig.\ \ref{FIG:dose} for low
($2.5\times10^{12}$\,cm$^{-2}$ at 1\,MeV,
 $1\times10^{13}$\,cm$^{-2}$ at 2.4\,MeV) and high ($1.25\times10^{15}$\,cm$^{-2}$ at 1\,MeV,
 $5\times10^{15}$\,cm$^{-2}$ at 2.4\,MeV) implant doses. At higher implant doses we observe
 that the THz pulse duration is shorter, and that the negative peak after
 the principal peak has both larger magnitude and smaller period (Fig.\ \ref{FIG:dose}(a)). In Fig.\ \ref{FIG:dose}(b) the Fourier
 transforms of the time-domain electric
 field show that at greater ion doses THz emission shifts to higher frequencies.

\begin{figure}[tb]
    \centering
    \includegraphics[width=6cm]{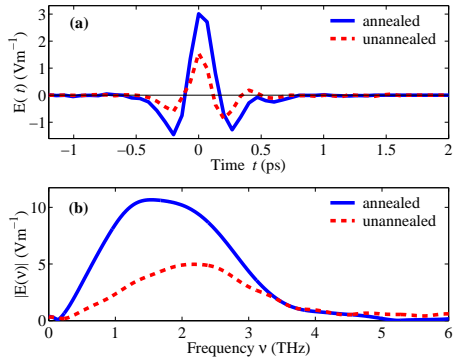}
    \caption{(Colour online) THz emission from unannealed (dotted lines) and annealed GaAs:As$^+$ (solid lines) for the lowest
 implant dose ($1\times 10^{13}$\,cm$^{-2}$ at 2.4\,MeV and $2.5\times 10^{12}$\,cm$^{-2}$ at 1\,MeV).
 Annealing causes an increase in maximum THz field from
 1.6\,Vm$^{-1}$ (unannealed) to 3.0\,Vm$^{-1}$ (annealed),
 but a decrease in the frequency of peak power from 2.1\,THz (unannealed) to 1.6\,THz (annealed).}
\label{FIG:anneal}
\end{figure}

The post-implant thermal annealing step (500{\degr}C, 30\,min.) was observed to result in larger
 peak THz fields, and greater power at low frequencies, as shown in Fig.\ \ref{FIG:anneal} for the lowest
 dose sample ($n_{\rm{d}}=2.5\times 10^{12}$\,cm$^{-2}$ at 1\,MeV,
 $n_{\rm{d}}=2.5\times 10^{12}$\,cm$^{-2}$ at 2.4\,MeV).
Annealing removes defects, at least partly repairing the damaged crystal structure.
Because defects scatter carriers, and can also trap conduction band electrons, we can
qualitatively predict that annealing will increase the pulse width, decreasing the
relative THz emission at high frequencies.

\section{Simulation}\label{SEC:sim}
In order to model the ultra-fast carrier dynamics leading to terahertz emission from semiconductor
 surfaces in such a way that reproduces the experimentally observed features it is appropriate to
 employ a three-dimensional Monte Carlo simulation, such as presented in
 Ref.~\onlinecite{prb02}. This model included the interactions between extrinsic and photogenerated
 carriers in $\Gamma$ and $L$-valleys, plasmon and magnetoplasmon effects,
 and the dielectric-air interface, and has been used to distinguish between the diffusion (photo-Dember) and
 surface field mechanisms of charge separation in {InAs} and {GaAs}\cite{sst04,prb02}, and to provide a
 quantitative explanation of the enhancement of THz emission under a magnetic field\cite{prb02}.

In this section we present simulations of the THz emission from GaAs:As$^+$ using a version of the
 model of Ref.~\onlinecite{prb02}, extended to include the effects of the significant concentrations of
 vacancies and trapping defects. All simulations used the physical properties of GaAs as given in
 Table 1 of Ref.~\onlinecite{prb02}, and the laser parameters of Sec.~\ref{SEC:exp}. Defect concentrations were
 assumed to be uniform over the range of infrared absorption, as is approximately the case in Fig.\ \ref{FIG:profile}.
 Altering the charged impurity (donor) concentration was not found to significantly alter the spectral shape of
 THz emission.

\subsection{Neutral impurity scattering}\label{SEC:vacScatt}
When As$^+$ ions collide with atoms in GaAs both the incident ion and the recoiling target atoms introduce
 vacancies, which act as carrier scattering centres. Calculations using SRIM\cite{SRIM} suggest that approximately 10,000
 vacancies are produced per incident ion, with the damage profile of Fig.\ref{FIG:profile}. For our unannealed samples
 this corresponds to vacancy concentrations ranging from $N_{\rm{vac}}\sim10^{19}$\,cm$^{-3}$
 to $\sim$10$^{22}$\,cm$^{-3}$. Electron paramagnetic resonance experiments have shown\cite{kaminska94} that
 the concentration
 of arsenic antisite defects is about 10$^{18}$\,cm$^{-3}$ after a 500\degr{C}, 30\,min.~anneal, for ion
 doses of $10^{16}$\,cm$^{-2}$. Since $N_{\rm{vac}}$ for our annealed samples therefore varies
 from 10$^{15}$\,cm$^{-3}$ to 10$^{18}$\,cm$^{-3}$, carrier-vacancy scattering cannot be disregarded as
 it normally is at lower defect concentrations.

Vacancies are included in the simulation by assuming that carriers scatter elastically from
 an electrically neutral spherical square well of width $R$, as defined by the potential $V = -V_0$ for
 carrier-vacancy distances $r \leq R$ and $V = 0$ for $r > R$. Following the partial-wave approach taken
 in Ref.~\onlinecite{ridley}, the momentum scattering rate $W_{\rm{vac}}$ can be calculated from the
 $l=0$ wave cross-section $\sigma$ using $W_{\rm{vac}}=\sigma N_{\rm{vac}} v$, where $v$ is the electron velocity
 before (and after) scattering. The resulting expression is:
\begin{equation}\label{EQN:vacscatt}
   W_{\rm{vac}} = \frac{4\pi\hbar N_{\rm{vac}}}{m^{*} k} \frac{(k \cot{kR} - \alpha \cot{\alpha R})^2}{(k^2+\alpha^2\cot^2{\alpha R})(1+\cot^2{kR})}
\end{equation}
\noindent where $\alpha$ is given by:
\begin{equation}\label{EQN:alpha}
  \alpha = \sqrt{\frac{2 m^{*} (E+V_0)}{\hbar^2}}
\end{equation}

The final state carrier is scattered uniformly into $4\pi$ steradians, because of the spherical
symmetry of the well. Taking the values of $R = 3$\,nm and $V_0 = 300$\,meV \cite{jap81_7295}
produces a scattering rate ranging from $\sim10^{11}$\,s$^{-1}$ at $N_{\rm{vac}} =
1\times10^{15}$\,cm$^{-3}$ to $\sim10^{14}$\,s$^{-1}$ at $N_{\rm{vac}} =
1\times10^{18}$\,cm$^{-3}$.

\begin{figure}[tb]
    \centering
    \includegraphics[width=6cm]{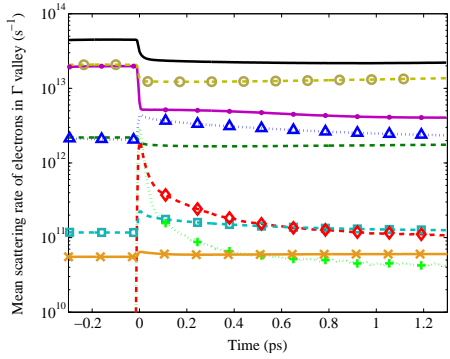}
    \caption{(Colour online)
Momentum scattering rates for electrons in the $\Gamma$-valley, with donor and vacancy
concentrations of $1\times10^{15}$cm$^{-3}$ each. The rate for each scattering mechanism is
averaged over all particles in the simulation, and is plotted as a function of time after laser
excitation. The largest contributions to the total rate (solid line) come from electron-hole
scattering (circles) and charged impurity scattering (dots). LO-phonon emission (triangles) and
absorption (dashed line) are also significant.  Acoustic phonon scattering (squares) and vacancy
scattering (diagonal crosses) produce lower rates. Least important are TO-phonon absorption
(diamonds) and emission (vertical crosses), where in both cases the electronic final state is in
the L-valley. The infrared pulse arrives at $t=0$\,ps, and produces a step in most scattering
rates.}
    \label{FIG:rates}
\end{figure}

To enable a comparison between these momentum scattering rates and those from mechanisms previously
included in the simulation, we plot in Fig.\ \ref{FIG:rates} the average momentum scattering rates
of electrons in the $\Gamma$-valley as a function of time $t$ after the infrared pulse for
$N_{\rm{d}},N_{\rm{vac}} = 1\times10^{15}$\,cm$^{-3}$. The scattering rates after $t=0$\,ps are
averaged over all photoinjected and extrinsic electrons, whereas before the infrared pulse arrives
only extrinsic electrons contribute. The time dependence of the scattering rates is a consequence
of the evolution of the carrier energy and concentration distributions. The greatest contributions
to the total rate are attributed to electron-hole scattering and charged impurities, which both
decrease at $t=0$ since they are inversely proportional to the local carrier density\cite{prb02}.
LO-phonon emission and absorption and TO-phonon emission and absorption followed by a jump into the
L-valley initially contribute $\sim$10\% of the total rate, but become less significant at later
times when the average carrier energy has decreased. Acoustic phonon and neutral vacancy scattering
provide only $\sim$1\% of the total rate.

\begin{figure}[tb]
    \centering
    \includegraphics[width=6cm]{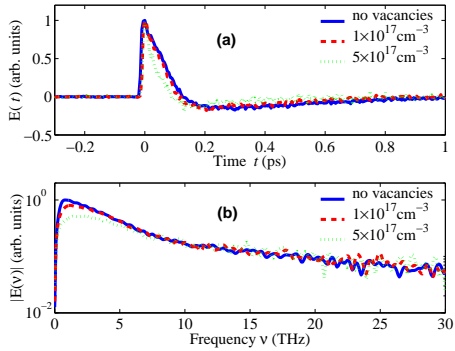}
    \caption{(Colour online) Simulated THz emission in (a) the time domain and (b) the frequency domain with no vacancies, and for
 vacancy concentrations of $1\times10^{17}$cm$^{-3}$, $5\times10^{17}$cm$^{-3}$. The FWHM of the spectra increase
 with vacancy concentration. Coherent phonon oscillations\cite{prl65_764} are not observed because the response
 of the crystal lattice is not included in the simulation. }
    \label{FIG:vacscatt}
\end{figure}

While Fig.\ \ref{FIG:rates} provides some insight into the significance of the various momentum
scattering mechanisms, the angular distribution of each mechanism must also be considered. At
$N_{\rm{vac}} = 1\times10^{17}$\,cm$^{-3}$ the vacancy scattering rate is
$\sim6\times10^{13}$\,s$^{-1}$, i.e.\ only 10\% of the total rate, yet as can be seen in Fig.\
\ref{FIG:vacscatt}(a) the THz pulse duration is reduced. This is a consequence of the uniform
angular distribution for vacancy scattering: the carrier direction is altered more significantly
than in the other mechanisms.

\subsection{Carrier-trapping defects}\label{SEC:trapping}
\begin{figure}[tb]
    \centering
    \includegraphics[width=6cm]{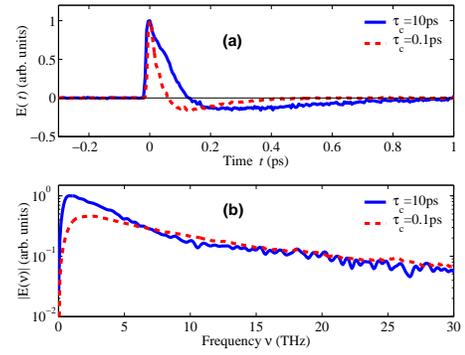}
    \caption{\label{FIG:tauc}(Colour online) Simulated THz electric field from bulk {GaAs} at carrier trapping times
 of $\tau_{\rm{c}} = 10$\,ps and
 $\tau_{\rm{c}} = 0.1$\,ps. In (a) the unfiltered simulated $\tau_{\rm{c}} = 0.1$\,ps pulse
 (dashed line) has shorter duration than
 the $10$\,ps (solid line). Their Fourier transforms are shown in (b).}
\end{figure}

The effect of trapping defects on THz emission was included in our model by introducing an exponential decay of the
 number of photoexcited carriers $n$ as a function of time $t$ after the infrared pulse according
 to the equation:
\begin{equation}\label{EQN:lifetime}
   n(t) = n(0) e^{-t/\tau_{\rm{c}}}
\end{equation}
\noindent where $\tau_{\rm{c}}$ is the carrier trapping lifetime. Time-resolved photoluminescence experiments
 of GaAs:As$^+$ implanted with a single $10^{16}$\,cm$^{-2}$ dose of 2\,MeV ions have measured the
 carrier trapping time to be as short as $\tau_{\rm{c}}=0.1$\,ps \cite{apl66_3304, apl76_1306}. It
 is important to
 distinguish $\tau_{\rm{c}}$ from the carrier recombination time, which is $\sim$4\,ps \cite{apl78_1667}. We assume
 that once trapped, carriers cannot escape the defect via thermal excitation, and that trapped carriers do not
 alter free carrier states. When the number of carriers $n$ is reduced on
 time scales of $<1$\,ps the electric field decay time after the pulse can be shortened, as shown
 in Fig.\ \ref{FIG:tauc}, resulting in a greater bandwidth. At $\tau_{\rm{c}}=0.1$\,ps
 the FWHM from the simulation is $7.8$\,THz, nearly 80\% larger than at a carrier
 trapping time of 100\,ps (FWHM $=4.4$\,THz).

\section{Comparison of experiment and simulation}
\label{SEC:expvsim}
\begin{figure}[tb]
    \centering
    \includegraphics[width=6cm]{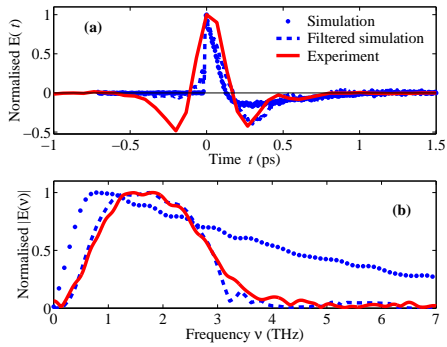}
    \caption{(Colour online) Comparison of simulated THz emission from GaAs with n-type
 doping $N_{\rm{d}} = 1\times10^{15}$\,cm$^{-3}$ using the
 low-dose sample data from Fig.\ \ref{FIG:dose}. In (a) the simulated electric field (dots) is seen
 to have a sharp rise $\sim10$\,fs in duration, and to drop sharply after $\sim0.1$\,ps, producing the broad spectrum
 in (b). When low-pass and high-pass filters are applied (taking account of the dispersion in ZnTe, and
 diffraction-limited collection respectively, as described in the text), the filtered spectrum
 (dashed line) can be seen to match the experimental spectrum (solid) extremely well. When
 transformed back into the time-domain, the filtered simulated electric field (dashed line of (a)) matches
 for times $\gtrsim0.2$\,ps.}
    \label{FIG:sim}
\end{figure}

The spectrum of the simulated THz emission has a larger magnitude at
 both low and high frequencies than the experimental data (Fig.\ \ref{FIG:sim}(b)). This can be accounted for
 by the effect of the EOS system and the parabolic mirrors used to collect the THz radiation. A
 frequency-domain picture of EOS with {ZnTe} has been presented by
 Gallot \emph{et.~al.~} \cite{apl74_3450,josab16_1204} that models
 the effect of the detection apparatus using three complex filters, namely: the spectrum of the autocorrelation of
 infrared pulse (which has negligible effect), the frequency-dependent electro-optic
 susceptibility $\chi^{(2)}$ of ZnTe, and the mismatch between the infrared
 group velocity and THz phase velocity in ZnTe. The resulting low-pass filter
 causes the high frequency components to reduce in magnitude, as shown in Fig.\ \ref{FIG:sim}(b). While the
 slow drop-off in simulated electric field at increasing frequencies suggests
 that large bandwidths are possible from surface field THz emitters, experimentally a thin ($\sim10\mu$m) ZnTe
 crystal is necessary to observe such frequency components, which would result in a prohibitively small signal.

While the experiment was designed to collect THz radiation with high efficiency, low frequency THz radiation emitted
 from the sample surface will be
 diffracted beyond the collection capability of the parabolic mirrors. By treating the two parabolic mirrors
 as a single thin lens, C\^{o}t\'{e} \emph{et.~al.} have calculated the high-pass filter
 corresponding to this effect\cite{josab20_1374}. They found that the approximate filter
 function $F_{\rm{high}}(\omega)=\textrm{erf}^{2}(2 \omega R \sigma / cf)$, where
 the THz radiation has a Gaussian beam-waist $\sigma$, and is collected by two parabolic mirrors each of
 focal length $f$ and radius $R$. For our experimental geometry ($f=15$\,cm, $R=2.5$\,cm, $\sigma=0.16$\,mm) this filter
 has a steep rise from 0 to 1 between 0 and $\sim1.5$\,THz, which when applied to the simulated data causes a reduction
 in low frequency components of the electric field.

Multiplying the simulated spectrum by these filters in the frequency domain produces a high
 accuracy match to the experimental
 spectra (Fig.\ \ref{FIG:sim}(b)). Taking the inverse Fourier transform of the filtered simulated spectrum results in a
 time-domain trace that compares adequately with the experimental time-domain data (Fig.\ \ref{FIG:sim}(a)). The
 low-pass (ZnTe) filter causes an increase both in the oscillation period and in the negative peak amplitude
 after the pulse. The dip in experimental electric field before the pulse is not reproduced
 by the application of the filters to the simulated data.

\begin{figure}[!b]
    \centering
    \includegraphics[width=6cm]{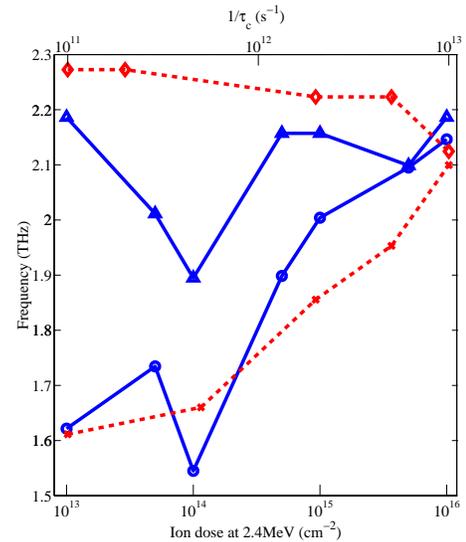}
    \caption{\label{FIG:fpeak}(Colour online) Frequency of peak emitted THz power $f_{\rm{p}}$ (circles, solid line) and FWHM of
 spectra (triangles, solid line) as a function of ion dose at 2.4\,MeV (the 1\,MeV ion dose
 was a quarter of the 2.4\,MeV dose).  $f_{\rm{p}}$ is defined as the midpoint of the two
 frequencies at which the amplitude of the Fourier transform of the electric field is 95\% of its maximum. The
 experimental data can be compared with values extracted from spectra produced by our carrier dynamics simulation
 (dashed lines), plotted as a function of inverse carrier
 trapping time $1/\tau_{\rm{c}}$. The trend in $f_{\rm{p}}$ (crosses) from the simulation reproduces the experiment. The
 FWHM from the simulation (diamonds) do not vary with $1/\tau_{\rm{c}}$, and are consistent with those measured.}
\end{figure}

The filters were applied to simulated spectra at carrier trapping times varying from $\tau_{\rm{c}}
= 0.1$\,ps to $\tau_{\rm{c}} = 100$\,ps. As shown in Fig.\ \ref{FIG:fpeak}, this enables a direct
comparison between the experimental $f_{\rm{peak}}$ as a function of ion dose, and $f_{\rm{peak}}$
extracted from the model as a function of $1/\tau_{\rm{c}}$. When $\tau_{\rm{c}} = 0.1$\,ps the
 filtered spectral peak is at $2.1$\,THz, in excellent agreement with the value ($2.2$\,THz) of
 the highest dose sample ($2.5\times10^{15}$\,cm$^{-2}$ at 1\,MeV, $1\times10^{16}$\,cm$^{-2}$ at 2.4\,MeV). As
 the carrier trapping time is increased
 ($1/\tau_{\rm{c}}$ decreased) the simulation tends to $f_{\rm{peak}} = 1.6$\,THz, in
 agreement with the experimental
 value. Values of the FWHM of filtered simulated spectra are approximately independent of
 $\tau_{\rm{c}}$, owing to the bandwidth limitations imposed by the $0.2$\,mm-thick ZnTe crystal.

A smaller increase in $f_{\rm{peak}}$ was observed over the experimental range of vacancy concentrations: at
 10$^{15}$\,cm$^{-3}$ $f_{\rm{peak}}=1.6$\,THz, and at 10$^{18}$\,cm$^{-3}$ $f_{\rm{peak}}=1.9$\,THz. This
 suggests that carrier trapping may be more significant than carrier-vacancy momentum scattering in determining
 THz pulse duration.

The experimental reduction in the spectral peak of THz emission after annealing (Fig.\ \ref{FIG:anneal})
 may be attributed to a combination of a reduction in carrier-vacancy scattering and an increase in carrier trapping
 time from $\tau_{\rm{c}} \sim 0.1$\,ps to $\tau_{\rm{c}} \sim 1$\,ps.

\section{Conclusion}\label{SEC:conc}
We have studied the surface field THz emission from GaAs implanted with multiple doses
 of high energy arsenic ions both experimentally and via a three-dimensional carrier dynamics simulation. The
 effect of trapping defects and neutral impurities on carrier dynamics was included in the model. At a short
 carrier trapping time of $0.1$\,ps our model predicts an increase in the FWHM of THz emission of
 nearly 80\% above the FWHM at long ($>10$\,ps) trapping times. This
 bandwidth increase was experimentally observed as a shift in peak spectral power towards higher
 frequencies at greater ion implant doses. We anticipate similar bandwidth improvements for photoconductive switches
 made from our GaAs:As$^+$ samples, but with greater available THz power.
The momentum scattering rate of carriers from vacancy defects was found to be $\sim10^{14}$s$^{-1}$, comparable
 to that of the two dominant scattering mechanisms in unimplanted GaAs (electron-hole and carrier-charged
 impurity scattering), for the vacancy concentrations expected in the highest dose sample. While the
 THz pulse duration decreased at high vacancy concentration, carrier-vacancy scattering was not found to account
 for the full shift in peak experimental spectral power.
While the simulation parameters were chosen to correspond to our GaAs:As$^+$ samples, the model is
 applicable to LT-GaAs, in which arsenic antisites (As$_{\rm{Ga}}$) are also the dominant defect, and
 which has comparable trapping times.

\section{Acknowledgements}\label{SEC:ack}
We would like to thank the EPSRC and the Royal Society for funding this research. E.C. thanks CONACyT (Mexico) for a scholarship. M.D.F.\ and C.J.\ acknowledge financial support from the Australian Research Council. Dr H.~Tan is gratefully thanked for help with the annealing experiments.



\end{document}